\documentclass[manuscript,nonacm] {acmart}
\usepackage{enumitem}
\usepackage{fancyhdr}
\AtBeginDocument{%
  \providecommand\BibTeX{{%
    \normalfont B\kern-0.5em{\scshape i\kern-0.25em b}\kern-0.8em\TeX}}}

\setcopyright{none}
\copyrightyear{2022}
\acmYear{2022}
\acmDOI{XXXXXXX.XXXXXXX}

\acmConference[CHI TRAIT '22]{Make sure to enter the correct
  conference title from your rights confirmation emai}{April 30,
  2022}{New Orleans, LA}
\begin{document}

\title{Exploring How Anomalous Model Input and Output Alerts Affect Decision-Making in Healthcare}

\author{Marissa Radensky}
\email{radensky@cs.washington.edu}
\orcid{0000-0002-5045-8269}
\authornote{Work done as intern at Microsoft and PhD student at University of Washington.}
\affiliation{%
  \institution{University of Washington}
  \city{Seattle}
  \state{WA}
  \country{USA}
}
\author{Dustin Burson}
\email{duburson@microsoft.com}
\orcid{}
\affiliation{%
  \institution{Microsoft}
  \city{Redmond}
  \state{WA}
  \country{USA}
}
\author{Rajya Bhaiya}
\email{Rajya.Bhaiya@microsoft.com}
\orcid{0000-0002-1473-5978}
\affiliation{%
  \institution{Microsoft}
  \city{Redmond}
  \state{WA}
  \country{USA}
}
\author{Daniel S. Weld}
\email{weld@cs.washington.edu}
\orcid{0000-0002-3255-0109}
\affiliation{%
  \institution{University of Washington \& Allen Institute for AI}
  \city{Seattle}
  \state{WA}
  \country{USA}
}


\begin{abstract}
  An important goal in the field of human-AI interaction is to help users more appropriately trust AI systems' decisions. A situation in which the user may particularly benefit from more appropriate trust is when the AI receives anomalous input or provides anomalous output. To the best of our knowledge, this is the first work towards understanding how anomaly alerts may contribute to appropriate trust of AI. In a formative mixed-methods study with 4 radiologists and 4 other physicians, we explore how AI alerts for anomalous input, very high and low confidence, and anomalous saliency-map explanations affect users' experience with mockups of an AI clinical decision support system (CDSS) for evaluating chest x-rays for pneumonia. We find evidence suggesting that the four anomaly alerts are desired by non-radiologists, and the high-confidence alerts are desired by both radiologists and non-radiologists. In a follow-up user study, we investigate how high- and low-confidence alerts affect the accuracy and thus appropriate trust of 33 radiologists working with AI CDSS mockups. We observe that these alerts do not improve users' accuracy or experience and discuss potential reasons why.
\end{abstract}

\ccsdesc[500]{Human-centered computing~Empirical studies in HCI}
\ccsdesc[500]{Computing methodologies~Machine learning}
\ccsdesc[500]{Applied computing~Health care information systems}

\keywords{human-AI interaction, explainable AI, appropriate trust of AI}

\maketitle

\section{Introduction and Related Work}
Human-AI decision-making teams are present in several domains such as loan risk assessment \cite{chromik2021think,green2019principles} and student performance forecasting \cite{dietvorst2015algorithm,weerts2019human}, but how to best help users appropriately trust AI models remains an open question broadly \cite{bansal2021does,zhang2020effect} and in healthcare specifically \cite{jacobs2021machine,naiseh2020explainability,micocci2021gps,gaube2021ai}. In healthcare, alert fatigue \cite{CashJaredJ2009Af,ancker2017effects} and high-stakes time constraints may further complicate achieving this goal.
Human-AI interaction research has given significant attention to healthcare \cite{holzinger2017we, lee2021human, cai2021onboarding, tonekaboni2019clinicians, antoniadi2021current, cai2019hello,yang2019unremarkable, wang2019designing, jacobs2021designing, hasling1984strategic, barnett1987dxplain}. Though a couple recent papers challenge explainable AI's utility in healthcare \cite{ghassemi2021false, duran2021afraid}, explainable AI clinical decision support systems (CDSSs) are studied in many areas such as antibiotic treatment \cite{lamy2020explainable}, antidepressant recommendation \cite{jacobs2021machine}, and acute critical illness prediction \cite{lauritsen2020explainable}. One common area is chest radiography, often utilizing saliency-map explanations \cite{murphy2020covid,ahsan2020study,karim2020deepcovidexplainer,teixeira2021impact,kim2021xprotonet,lanfredi2021comparing,gerlings2021explainable}. We should note, nevertheless, that concerns exist around the faithfulness and utility of saliency-map explanations \cite{reyes2020interpretability, jain2019attention,viviano2019saliency}.
Regarding AI's future impact on radiologists, hopes \cite{van2019survey,waymel2019impact,ooi2021attitudes,jha2016adapting} may be buttressed and concerns \cite{van2019survey,mazurowski2019artificial} mitigated through efforts to understand and improve radiologist-AI teams. For example, Xie et al. discovered many insights around desired explanations for a chest x-ray CDSS \cite{xie2020chexplain}. Meanwhile, providing diagnostic advice for chest x-rays, Gaube et al. observed that radiologists more so than other physicians thought human expert suggestions were lower quality when presented as AI suggestions. They also found that incorrect suggestions led to decreased accuracy, regardless of the alleged suggester \cite{gaube2021ai}.

Model explanations are often sought due to perceived anomalous behavior \cite{gregor1999explanations}.
Tomsett et al. explained that communicating uncertainty due to \textbf{anomalous model input} is important, as it helps users to construct a more accurate mental model of the AI \cite{tomsett2020rapid}. Many works seek to detect out-of-distribution input \cite{fariha2021conformance,ruff2018deep,devries2018learning,ren2019likelihood,yang2021generalized}, and some focus on chest x-rays in particular \cite{cohen2019chester,cao2020benchmark,zhang2020viral,berger2021confidence}. Multiple works have explored how users may interact with out-of-distribution data \cite{liznerski2020explainable,mokoena2022anomaly,chen2020oodanalyzer,delaney2021uncertainty,liu2021understanding}. Most related to our work, Suresh et al. saw that providing information about a model's training data did not improve user accuracy when the input image was abnormal, but participants were less likely to follow incorrect recommendations for such input \cite{suresh2020misplaced}. With respect to \textbf{anomalous model confidence}, Suresh et al. observed that providing a model's predicted class probabilities did not impact accuracy when the probabilities were abnormal in the sense that they were low and similar, but participants were less likely to follow incorrect recommendations with such class probabilities \cite{suresh2020misplaced}. Also, Bussone et al. found that providing an AI CDSS's confidence as relatively high or low only slightly impacted healthcare professionals' trust of and reliance on the AI \cite{bussone2015role}. Regarding \textbf{anomalous model explanations}, some works have looked into determining when explanations have reduced reliability \cite{qiu2021resisting,merrick2020explanation}. In addition, DeGrave et al. saw saliency-map explanations for COVID-19 in chest radiographs that could be considered anomalous in that they depict spurious correlations developed in training \cite{degrave2021ai}, also seen in another work \cite{maguolo2021critic}.

Despite all this work related to anomalous model input and output, to the best of our knowledge, this is the first work to investigate how AI anomaly alerts may help users to more appropriately trust AI. We begin with a formative study in which radiologists and other physicians are interviewed and surveyed about their reactions to anomaly-alert mockups for a CDSS used to find pneumonia in chest x-rays. We flag anomalies for three main aspects of a model's communication with users. Two of the anomaly types are very high and low confidence with respect to the model's confidence distribution. Another is input significantly different from the model's training data, presented here as pediatric x-rays, which have been noted as a potential form of anomalous input for models evaluating chest radiographs \cite{garbin2021structured}. The last is explanations significantly different from an expected average explanation, presented here as saliency maps focusing outside the lungs, which prior work has observed \cite{degrave2021ai}. We follow up with a user study examining how the two confidence alerts impact radiologists' accuracy in working with mockups for the same CDSS. 

In summary, we make the following contributions:
\begin{itemize}
    \item a mixed-methods formative study with 4 radiologists and 4 other physicians exploring how users of a clinical decision support system for evaluating chest x-rays react to alerts for different anomalous model input and output: very high confidence, very low confidence, anomalous input, and anomalous explanations.
    \item a 33-participant user study investigating high- and low-confidence alerts' effect on radiologist-AI team accuracy.
    \item evidence suggesting that 1) the four proposed anomaly alerts are desired by non-radiologist physicians, and 2) high-confidence alerts are desired by both radiologists and non-radiologists, but 3) high- and low-confidence alerts are not necessarily helpful for improving radiologist-AI team accuracy.
\end{itemize}

\section{Study 1: Mixed-Methods Formative Study}
\subsection{Study Design}
\subsubsection{Research Questions}
Study 1's research questions were as follows: 1) how do users of a chest x-ray CDSS react to alerts for very high and low confidence, anomalous input, and anomalous explanations?, 2) do users think any of the alerts would help their decision-making?, and 3) how should the alerts be presented in order to be most useful?

\subsubsection{Example Selection and Presentation}
Each participant evaluated the same 26 chest x-rays, taken from the train set of the Kaggle RSNA Pneumonia Detection Challenge dataset \cite{shih2019augmenting}, which has labels carefully assigned by at least one expert radiologist. The x-rays were evaluated for pneumonia using the DenseNet121 model from the TorchXRayVision library \cite{cohen2021torchxrayvision,cohen2020limits} pre-trained on the same dataset. We used a threshold of 0.5 for determining each image's classification, providing a high sensitivity of 95.2\% at the cost of a low specificity of 57.9\%. A cardiac and neuro ICU charge registered nurse (CCRN, MSN) confirmed that all the examples were reasonably labeled for pneumonia. The saliency-map for each example was a Grad-CAM++ explanation \cite{chattopadhay2018grad} generated using publicly available PyTorch implementations \cite{simplepytorchcam,jacobgilpytorchcam}. 

Each participant interacted with all 16 anomaly scenarios, consisting of every combination of anomaly type (high or low confidence, explanation, or input), alert presence (alert or no alert), and prediction accuracy (correct or incorrect). The exception was that there were two low-confidence alert scenarios with incorrect predictions and none with a correct prediction. Each participant also saw ten non-anomalous scenarios, half of which had a "pneumonia" prediction. Of the non-anomalous scenarios, one "pneumonia" and one "no-pneumonia" prediction were incorrect. Two non-anomalous scenarios were dedicated to training, and both had correct predictions, one "pneumonia" and one "no-pneumonia." To avoid too many variables for the small sample size, the example for each scenario and the prediction for each anomaly type were held constant. The prediction for each anomaly type was chosen based on when its alert would likely be more useful ("pneumonia" for high confidence and anomalous explanations and "no pneumonia" otherwise).


The 26 chest x-rays were selected from the dataset to represent each of the aforementioned scenarios. Unlike the other examples, those labeled for \textbf{anomalous input} were selected to have large black borders and appear like pediatric cases, as confirmed with a cardiac and neuro ICU charge registered nurse (CCRN, MSN). Classifying pediatric cases is a potential issue for models trained on adult x-rays. This is noted, for example, in the chest-radiograph dataset CheXpert's datasheet \cite{garbin2021structured}. In order to have enough examples of incorrectly-classified anomalous input, one anomalous input that was classified correctly with 73\% confidence was instead presented as classified incorrectly with 60\% confidence. Furthermore, unlike the other examples, those labeled for \textbf{anomalous explanations} were selected to have a saliency-map explanation focused outside the lungs. This kind of anomalous explanation has been observed occurring in the real world as the result of spurious correlations \cite{degrave2021ai}. For a confidence range of 50\% to 100\%, x-rays with an associated AI confidence between 53\% and 55\% were selected for the \textbf{anomalously-low-confidence} scenarios, and x-rays with an associated confidence between 97\% and 99\% were selected for the \textbf{anomalously-high-confidence} scenarios.
The anomalous confidence levels were chosen to be on the extreme ends of the confidence range so that the results would generalize better to other models, whether or not their highest and lowest confidence levels tend to be as extreme. If a scenario was associated with a correct "pneumonia" prediction but not with an anomalous explanation, then the representative x-ray was selected such that its explanation mostly matched the dataset's ground-truth mask for pneumonia. To prevent bias against the AI based on initial interactions, sections of scenarios were created to prioritize showing correct predictions earlier as well as negative anomaly alerts later, as prior work shows that algorithmic aversion arises after observing an algorithm fail \cite{dietvorst2015algorithm}. The examples within each section were randomly ordered.

\subsubsection{Participants and Procedure}
\label{sec:proc1}
Eight participants were recruited through contacts and directories and donated their time to the study. Four are radiologists, while the other four are physicians in other areas. Two of the radiologists evaluate chest x-rays monthly, while the other participants evaluate chest x-rays weekly. The study sessions were conducted over Microsoft Teams and were around 50 minutes long. Each participant started with an initial Microsoft Forms survey of ten 7-point Likert-type questions to gauge their inclination towards using AI and familiarity with AI.\footnote{The full list of questions for this survey may be found at \href{https://docs.google.com/document/d/1Zt-zuPwUbY4-GgrEnY-xpCltHrqAD1kZ74HUBeadyLU/edit?usp=sharing}{this link}.} These questions were selected based on prior work \cite{abuzaid2022assessment,van2019survey,pinto2019medical,waymel2019impact,ooi2021attitudes,collado2018role}. Next, the participant was shown a Microsoft PowerPoint over Microsoft Teams.\footnote{An example of the slides seen by a given participant may be found at \href{https://docs.google.com/presentation/d/1Bv54JZgD1vciTyfG_O3Aduuf3Ij9R7rX/edit?usp=sharing&ouid=111594202971372741587&rtpof=true&sd=true}{this link}.} The participant was instructed on how to work with the CDSS mockups to evaluate chest x-rays for pneumonia. As recommended by Bussone et al. \cite{bussone2015role}, the confidence was defined and described as going from 50\% to 100\%, with 50\% indicating that the AI is completely unsure if the patient has pneumonia or not. The participant then evaluated two training examples followed by 24 more. For the training examples, the researcher told them the correct answer after they made their decision. Participants were asked to think aloud \cite{van1994think} while evaluating the x-rays, and the researcher asked occasional questions about their interaction as well as entered their final answers.


\begin{figure}[tb]
  \centering
  \fbox{\includegraphics[width=.22\linewidth,height=4.25cm]{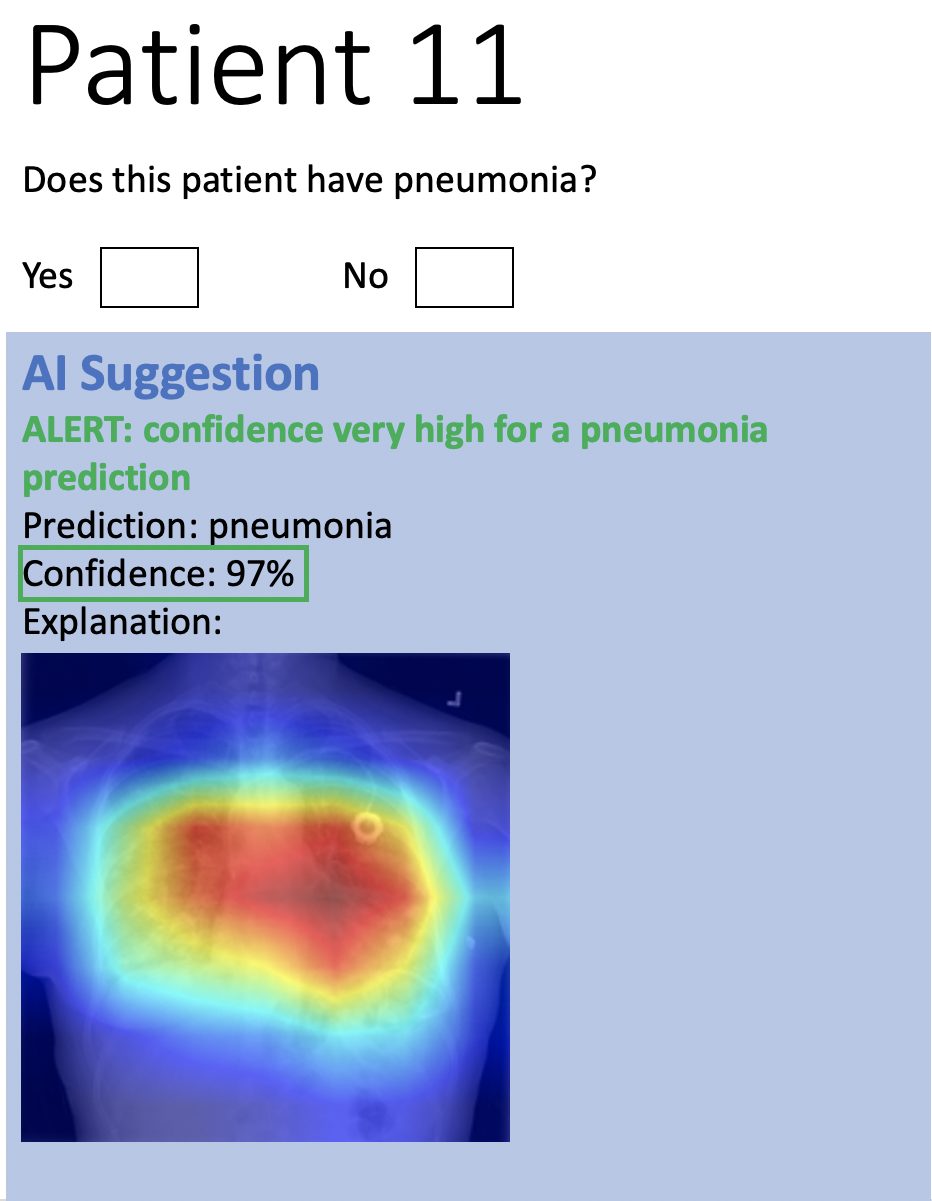}}
  \fbox{\includegraphics[width=.22\linewidth,height=4.25cm]{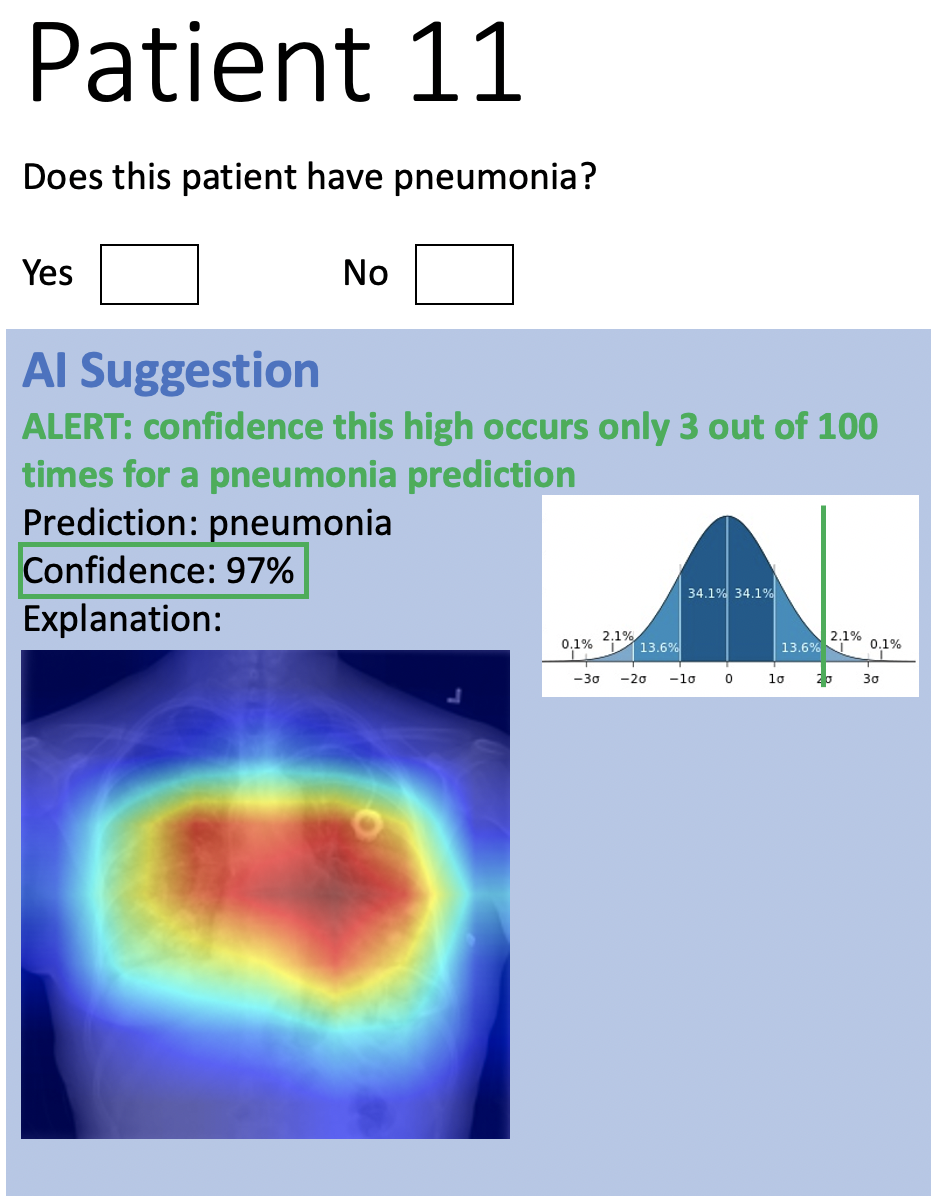}}
  \fbox{\includegraphics[width=.22\linewidth,height=4.25cm]{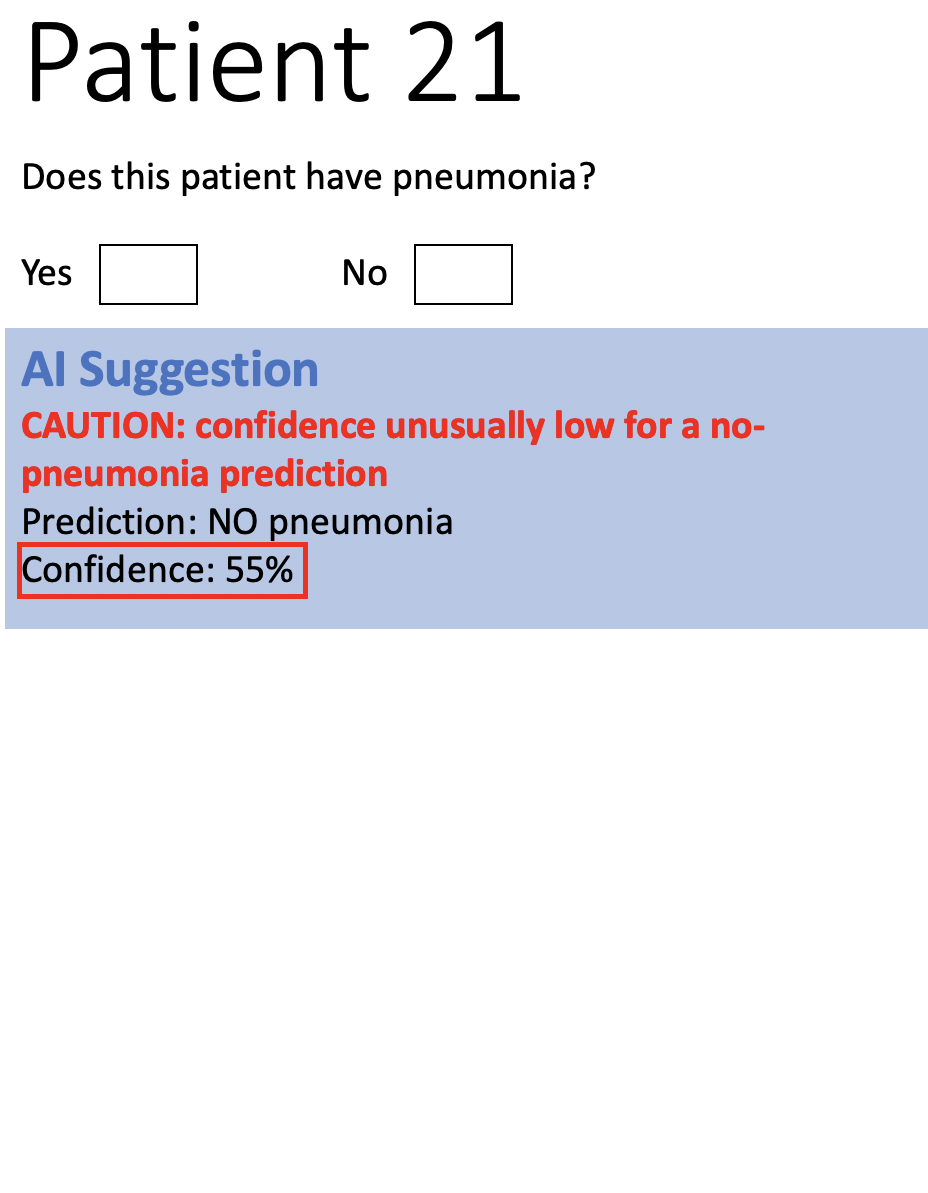}}
  \fbox{\includegraphics[width=.22\linewidth,height=4.25cm]{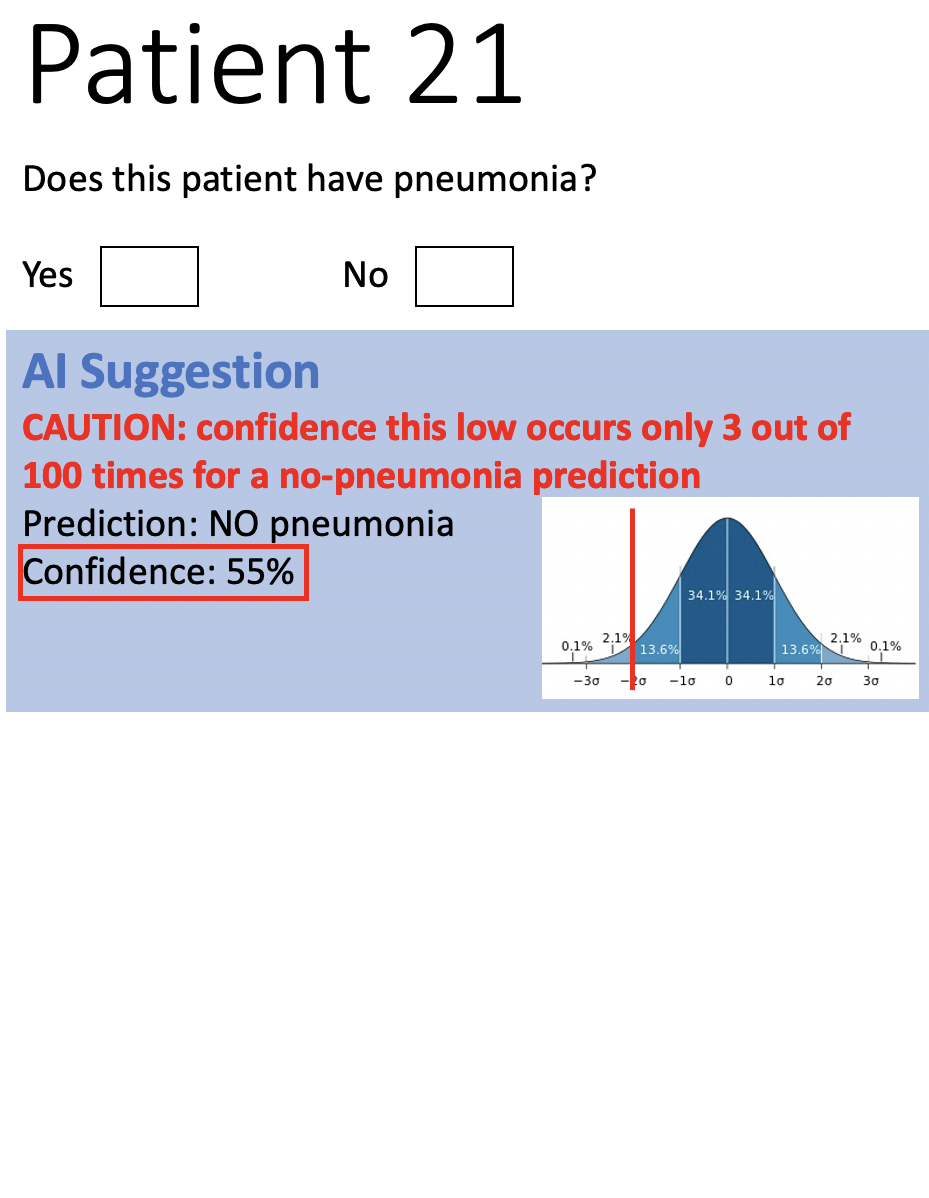}}
  \fbox{\includegraphics[width=.22\linewidth,height=4.25cm]{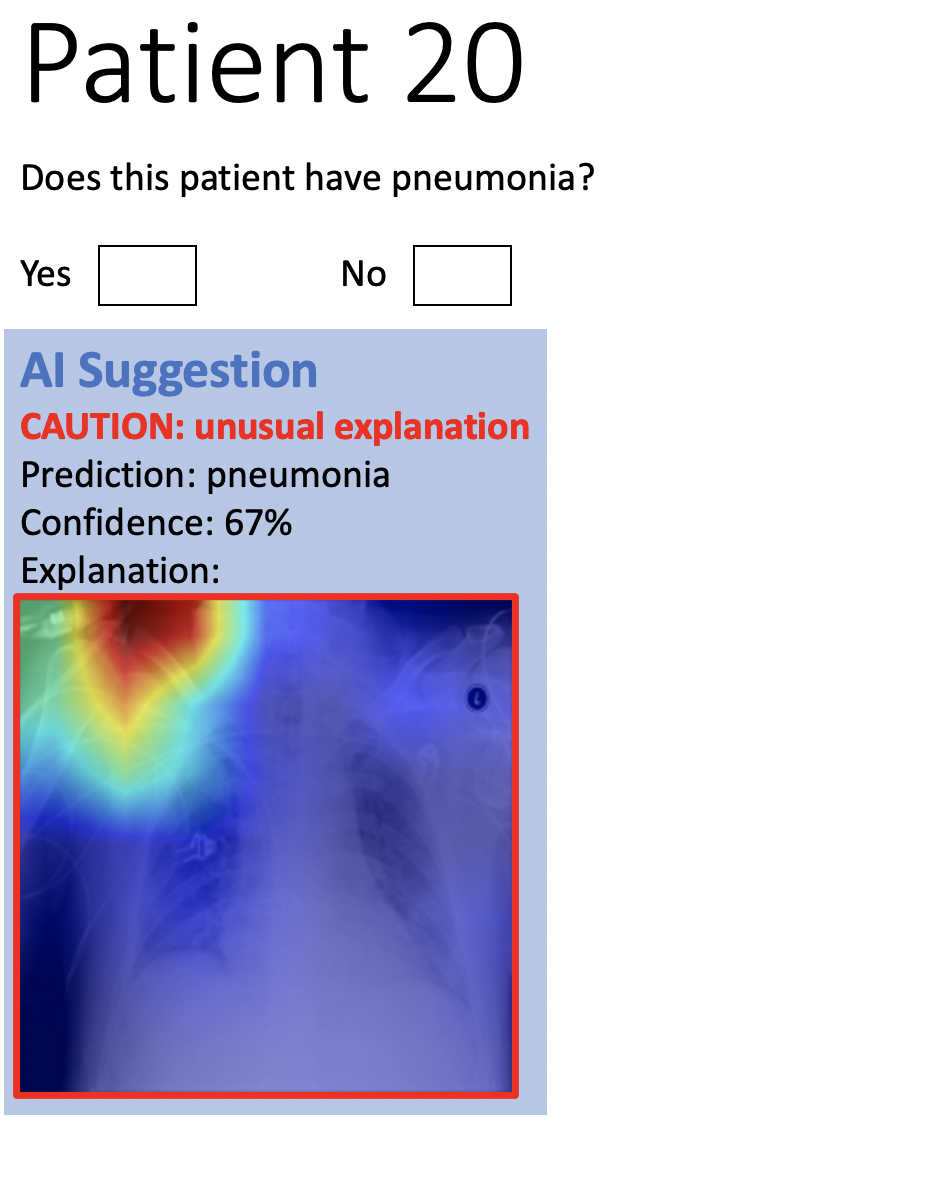}}
  \fbox{\includegraphics[width=.22\linewidth,height=4.25cm]{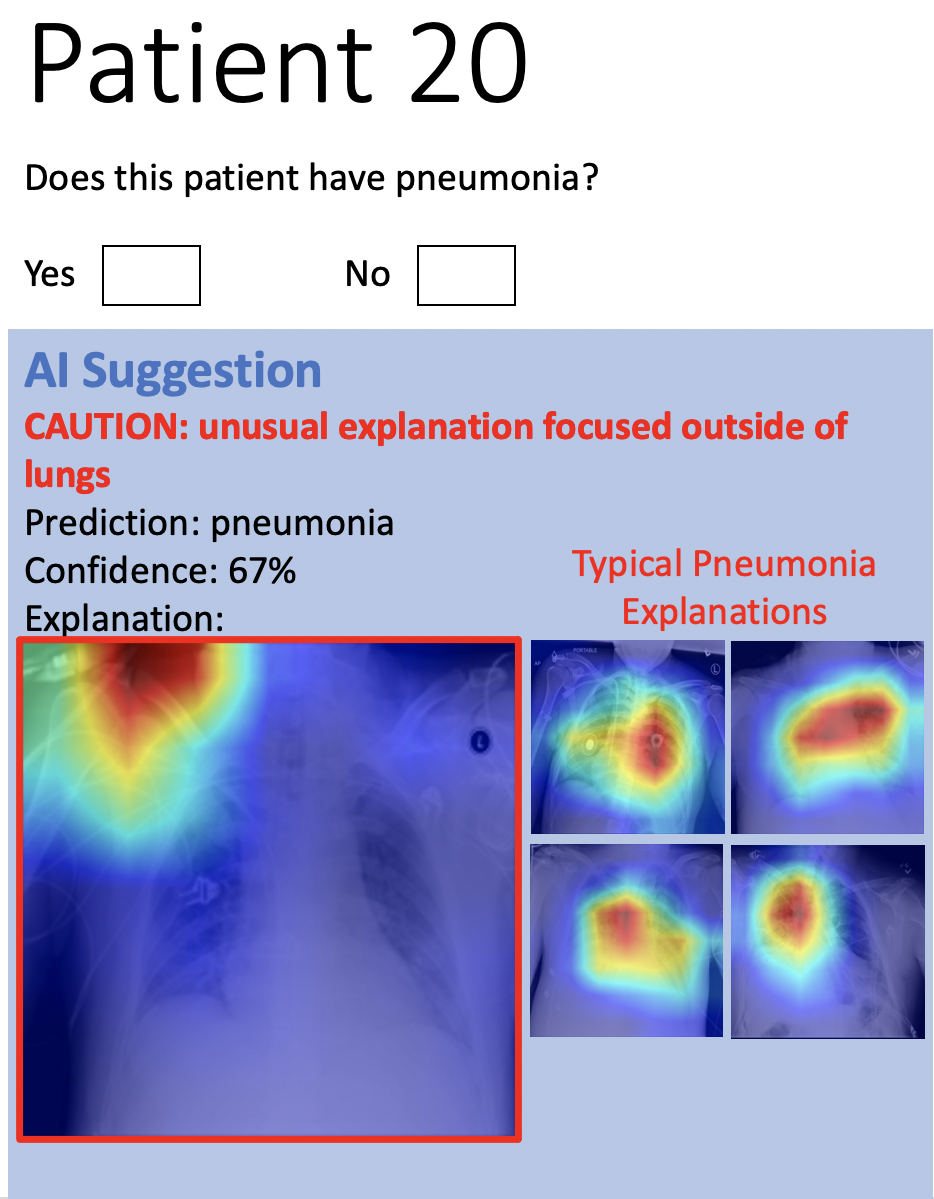}}
  \fbox{\includegraphics[width=.22\linewidth,height=4.25cm]{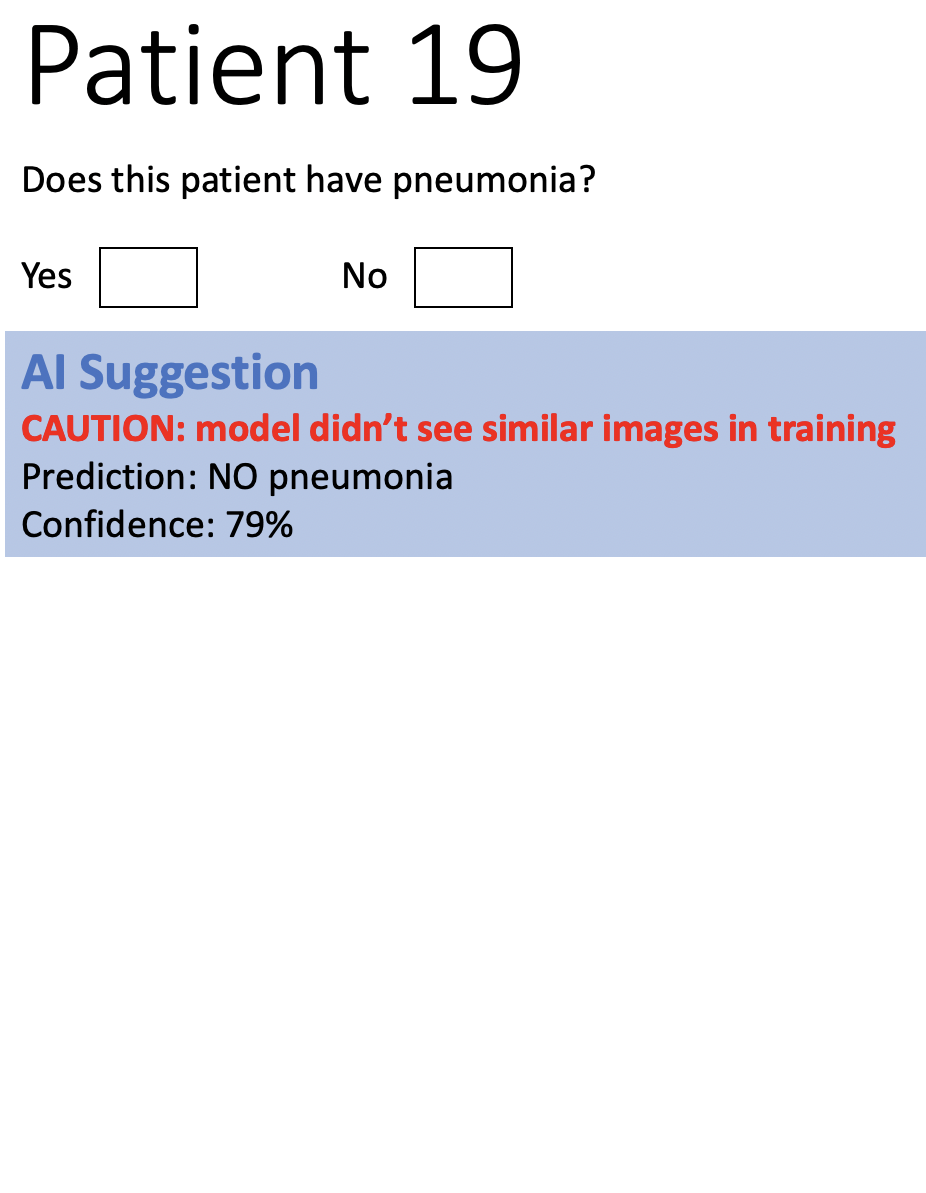}}
  \fbox{\includegraphics[width=.22\linewidth,height=4.25cm]{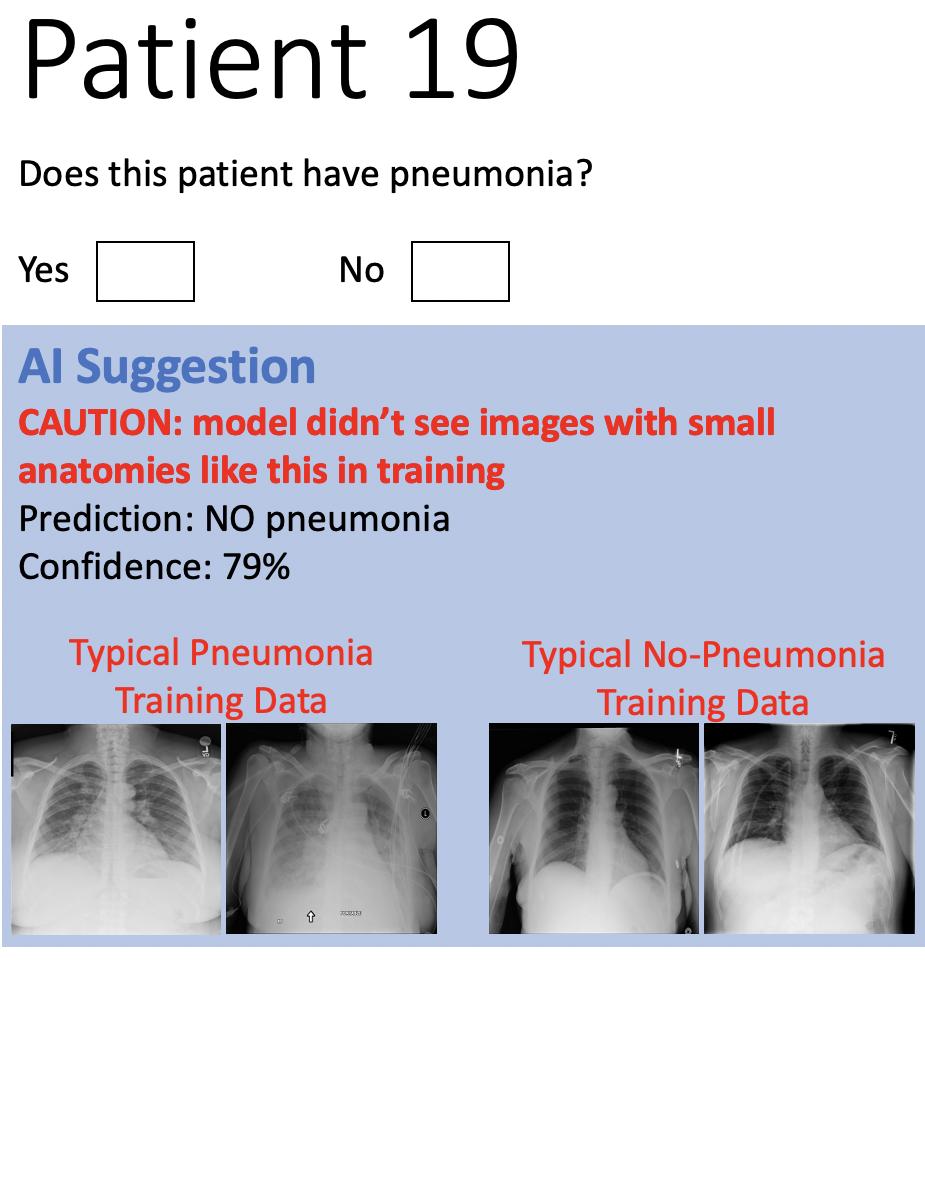}}
  \caption{Study 1 anomaly alert mockups. Top row: high confidence, high confidence detailed, low confidence, low confidence detailed. Bottom row: anomalous explanation, anomalous explanation detailed, anomalous input, anomalous input detailed.}
  \label{fig:anomalerts}
  \Description{Each alert is presented in the context of the system's header for each patient, such as "Patient 11," and the question asked of the participant: "Does this patient have pneumonia?", underneath which there is a "yes" box followed by a "no" box from which the participant may choose. The AI suggestion below is encompassed by a light blue box. Each alert puts a short phrase below the header "AI Suggestion," followed by the prediction (e.g., "Prediction: pneumonia"), confidence (e.g., "Confidence: 97\%"), and saliency-map explanation if predicting "pneumonia." For high confidence, the simple version says "ALERT: confidence very high for a pneumonia prediction." It places the confidence in a green-bordered box. The detailed version makes the alert phrase longer and says, "ALERT: confidence this high occurs only 3 out of 100 times for a pneumonia prediction." It shows a standard-distribution curve with a vertical green line through the point around 97\%. For low confidence, the simple version says "CAUTION: confidence unusually low for a no-pneumonia prediction." It places the confidence in a red-bordered box. The detailed version makes the alert phrase longer and says, "CAUTION: confidence this low occurs only 3 out of 100 times for a no-pneumonia prediction." It shows a standard-distribution curve with a vertical red line through the point around 55\%. For the anomalous explanation, the simple version says, "CAUTION: unusual explanation," and puts a red border around the explanation. The detailed version says, "CAUTION: unusual explanation focused outside of lungs." It shows 4 typical pneumonia explanations to the right of the main explanation. For the anomalous input, the simple version says, "CAUTION: model didn't see similar images in training, and the detailed version says, "CAUTION: model didn't see images with small anatomies like this in training." The detailed version also shows 2 instances of typical pneumonia and no-pneumonia training data.}
\end{figure}

Each x-ray was presented in a Microsoft PowerPoint slide alongside an AI suggestion, as shown in Figure \ref{fig:anomalerts}. The prediction and confidence were always provided, and the saliency-map explanation was included if the prediction was "pneumonia." Each anomaly alert provided a short phrase describing the anomaly and highlighted the anomalous information with a colored box - green for high confidence and red otherwise. For each alert, when it was shown in a more useful scenario (i.e., when the AI was correct for the high-confidence case and incorrect for the other cases), an additional slide provided a more detailed version of the alert with a longer phrase and additional information. The simple and detailed versions of each alert are presented in Figure \ref{fig:anomalerts}. Inspired by Xie et al. \cite{xie2020chexplain}, the anomalous-input detailed alert included two example images of typical pneumonia training data and no-pneumonia training data, and the anomalous-explanation detailed alert included four examples of typical pneumonia saliency-maps to contrast with the anomalous one. For the high-confidence and low-confidence detailed alerts, a bell curve with a line through the point where the confidence level was served as the model's imaginary confidence distribution. This curve was not based on the model's actual confidence distribution but was merely used to see how people would react to such a visualization.

After evaluating the x-rays, the participant engaged in a semi-structured interview\footnote{The full script for the study sessions may be found at \href{https://docs.google.com/document/d/1N_PWWx4Mkk23_Ds3RKHJS-4u8ToE9rDf/edit?usp=sharing&ouid=111594202971372741587&rtpof=true&sd=true}{this link}.} to discuss their general use of the AI, when they challenged or trusted it, when it affected their confidence, and when it affected their efficiency. These questions were primarily probes for investigating how the alerts impacted their experience. They then completed a final Microsoft Forms survey consisting of two 7-point Likert-type questions for each anomaly type: 1) "The X alerts improved how I used the AI suggestions," and 2) "If I use a similar system in the future, I would like to receive X alerts." The survey reminded participants of how each alert's simple and detailed format appeared. As time permitted, the interview was recommenced to directly discuss how each alert affected their experience and any additional questions.

\subsection{Results and Discussion}
\begin{figure}[tb]
  \centering
  \includegraphics[width=10.5cm]{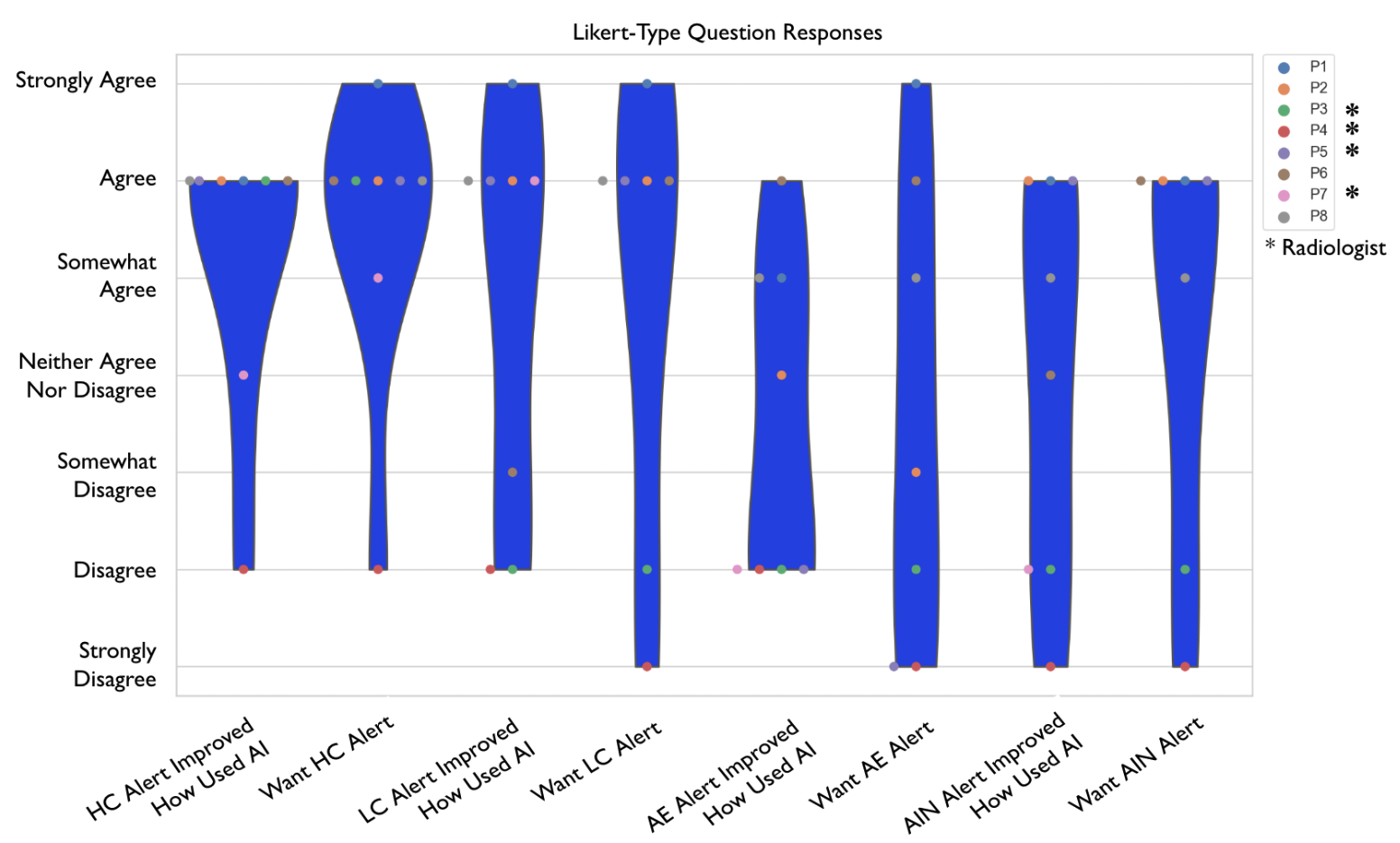}
  \vspace*{-.1in}
  \caption{Study 1 final survey responses. Color-coded dots represent individuals (see legend). Asterisks mark radiologists. We observe that HC alerts are overall desirable, and non-radiologist participants generally desire each alert. Note: P7 left Q4, Q6, and Q8 blank. }
  \label{fig:finalLikert}
  \Description{Eight violin plots are shown to correspond to the eight questions in Study 1's final survey. For improving how the participant used the AI suggestion, high-confidence alerts received 6 votes for "agree," 1 for "neither agree nor disagree," and 1 for "disagree." Low-confidence alerts received 1 vote for "strongly agree," 4 votes for "agree," 1 for "somewhat agree," and 2 for "disagree." Anomalous-explanation alerts received 1 vote for "agree," 2 votes for "somewhat agree," 1 vote for "neither agree nor disagree," and 4 votes for "disagree." Anomalous-input alerts received 3 votes for "agree," 1 for "somewhat agree," 1 for "neither agree nor disagree," 2 for "disagree," and 1 for "strongly disagree." For desirability, high-confidence alerts received 1 vote for "strongly agree," 5 votes for "agree," 1 vote for "somewhat agree," and 1 vote for "disagree." Low-confidence alerts received 1 vote for "strongly agree," 5 votes for "agree," 1 vote for "disagree," and 1 vote for "strongly disagree." Anomalous-explanation alerts received 1 vote for "strongly agree," 1 vote for "agree," 1 vote for "somewhat agree," 1 vote for "somewhat disagree," 1 vote for "disagree," and 2 votes for "strongly disagree." Anomalous-input alerts received 4 votes for "agree," 1 vote for "somewhat agree," 1 vote for "disagree," and 1 vote for "strongly disagree." The participants who gave at least two negative responses were the four radiologists.}
\end{figure}

The think aloud and interview responses were analyzed by the first author using thematic analysis \cite{braun2012thematic}. The Likert-type responses to the final survey are presented in Figure \ref{fig:finalLikert}; please note that P7 left three questions blank.

\subsubsection{High- and Low-Confidence Alerts}
\label{sec:LC}
\textbf{The high confidence (HC) alert was desired by all but one participant, and most participants thought it improved their AI use.} This makes sense given that it is the only positive alert indicating that the AI suggestion should be more useful. However, most participants found the detailed version of the alert no more helpful than the simple one. P4 remarked, "\emph{Confidence of 97\% said the same thing to me}," implying that the additional information was unclear in showing how \emph{common} the confidence of 97\% was. \textbf{The HC alert was primarily described as helpful for reconsidering one's initial answer when disagreeing with the AI.} Four participants saw the HC alert as useful in this way. P8 explained, “\emph{[The HC alert] mainly made me think more in situations where I disagreed with it}.” In particular, a couple participants found the HC alert useful, when disagreeing with the AI, for indicating the need to seek more information, whether in the form of another opinion or additional patient details. P2 noted, “\emph{In that [HC alert] situation, if you disagree with the second observer, you want a third observer}," while P3 reflected, "\emph{I think in that situation [in which I might disagree with a HC alert], I would have then sought… more history, previous images, etc}.” On the other hand, a couple participants expressed concern that the alert might bias them towards the AI's answer. P2 commented, "\emph{It's almost biasing me, trying to push me down the pneumonia route}." However, P1 noted that this may not always be concerning with respect to pneumonia predictions. They explained, “\emph{It's always better to have to overcall things and have a slightly higher false positive rate}." Thus, HC alerts biasing users towards false-positive concern for pneumonia may be more acceptable than those biasing users towards false-negative lack of concern for pneumonia.

\textbf{All the non-radiologists desired the low confidence (LC) alert, and most of them thought it improved their AI use. However, the radiologists were divided on its utility and desirability.} This may be because non-radiologists rely more on the AI and thus find it more useful to recognize decreased reliability. As with the HC alerts, participants largely found the detailed version unnecessary. P2 noted, "\emph{I think it'll confuse the average clinician}." \textbf{The most commonly cited use for the LC alert was to recognize that one may disregard the AI suggestion.} Three radiologists described the LC alert as such. P5 commented, "\emph{I think it made me just kind of rely on my own skills}."

\subsubsection{Anomalous-Explanation Alerts}
\textbf{The anomalous-explanation (AE) alert was the least popular alert but more popular with non-radiologists.} Three non-radiologists rated the alert as desirable and having improved their AI use. As mentioned earlier, non-radiologists may like the alert more because they gain more from the AI. Regarding the detailed version, three participants acknowledged its extra words as helpful. P8 explained, “\emph{It's clearly focused on a point outside the lung, but I think having that detail in case there was a different reason why there was something unusual is helpful}.” However, two participants noted disliking the additional words, and another two remained confused by the alert's meaning. P4 recounted, "\emph{The words were confusing because it implied that this was some exceptional case where we actually have some brilliant conclusion}." The alert's wording thus requires careful iteration if it is to be utilized. To start, the alert should clarify its association with \emph{reduced} reliability. As for the detail of showing typical pneumonia explanations, all but one participant either found it unhelpful or was confused by it. 

\textbf{While participants often recognized the anomalous explanations without an alert, a few participants still found the AE alert helpful for improving their understanding or trust of the AI.} Participants commonly disregarded an anomalous explanation with or without the alert. For example, with no alert, P7 correctly perceived, "\emph{The heatmap is highlighting the neck and the head? That literally makes no sense. I think there could be pneumonia for this one, but it's not where the heat map is}.” However, three non-radiologists still described the alert as helpful. For instance, P6 noted how it helped them build their understanding of the AI: "\emph{I think that [the AE alert] also is helpful because then… you know that it's... either outside the training model, or it's picking up something that's unexpected.}" Meanwhile, the alert increased P8's trust of the AI. They said, “\emph{It improved my confidence in the AI's abilities…. It is at least identifying that the heatmap is probably in the wrong area, and that this is not a good image or it is not well-trained in this image.}" 

\subsubsection{Anomalous-Input Alerts}
\textbf{The anomalous-input (AIN) alert was more popular with non-radiologists in terms of desirability and improving AI use.} All non-radiologists rated the alert as desirable and three rated it as improving AI use. In contrast, only one radiologist viewed the AIN alert as desirable and improving AI use. As discussed previously, non-radiologists may appreciate the alert more because they rely more on the AI. Furthermore, the alert's detailed version was overall considered unhelpful. Though no participant described the typical training data as useful, three noted that it might be helpful in other contexts such as for detecting rarer diseases or for assisting junior doctors. 

\textbf{In response to the AIN alert, participants often ignored the AI suggestion, but participants were divided on whether or not the alert improved system use.} Four participants noted disregarding the AI due to the alert. P6 explained, "\emph{It just made me sort of discount what the AI had to say in that situation because it seemed like it was kind of outside the training model}." Interestingly, the three most familiar with AI (P1, P2, P5) were most positive about the alert in their Likert-type responses. P2 commented, "\emph{It's obvious the AI model hasn't seen as many kids, so that's why that caution has come up, which I think is quite useful because it's kind of telling the human… we haven't looked at as many inferences as we should do, and therefore our level of confidence is lower}." Conversely, a couple participants found the AIN alert unhelpful and even irritating. P3 explained, “\emph{If AI is making an excuse for itself, then it kind of is more of a nuisance than anything}." Likewise, P4 insisted, "\emph{The confidence really says everything that I need to know}." Meanwhile, P7 expressed confusion: "\emph{I just had to read it like multiple times to figure out exactly what it was trying to say}." 
Thus, the AIN alert may be improved with a recommendation to re-train the model with similar data.

\subsubsection{AI Suggestions with Reduced Reliability}
When asked about when the AI may have slowed them, three participants noted at least one alert. Decreasing the information provided when the AI has reduced reliability may ameliorate this issue. We asked the last six participants if they would 1) want an AI suggestion when it had reduced reliability and 2) want the AI to specify what kind of anomalous behavior (LC, AE, AIN) contributed to reduced reliability. We obtained mixed results. Two participants did not want an AI suggestion when there was reduced reliability, while the others still wanted a suggestion in at least one anomalous situation. In addition, one preferred that the AI not specify why it has reduced reliability, one wanted simply to judge the AI's reliability based on the saliency map, one was unclear, and the rest wanted specification for at least one anomaly type. Future work may investigate these questions further. Prior work indicates that AI suggestions of all confidence levels should be provided to help users calibrate their mental models of the AI \cite{bussone2015role}, but systems have been designed to avoid providing AI suggestions when faced with anomalous input \cite{cohen2019chester}.

\section{Study 2: User Study}
\subsection{Study Design}
\subsubsection{Research Question and Hypotheses}
\label{sec:hypo2}
Study 2's research question was: do alerts for very high and low AI confidence improve human-AI team accuracy in the context of a CDSS for evaluating chest x-rays? We focused on these alerts because: 1) they were most popular with Study 1's radiologists, and 2) the others seemingly required significant design iteration. Our hypotheses state that human-AI team accuracy is \textbf{H1}: \underline{affected} by \underline{high}-confidence alerts when the AI is \underline{correct}, \textbf{H2}: \underline{not affected} by \underline{high}-confidence alerts when the AI is \underline{incorrect}, \textbf{H3}: \underline{affected} by \underline{low}-confidence alerts when the AI is \underline{incorrect}, and \textbf{H4}: \underline{not affected} by \underline{low}-confidence alerts when the AI is \underline{correct}.

\subsubsection{Example Labeling}
\label{sec:label}
Because this study was quantitative, we wanted to make sure that the example x-rays used had strongly reliable labels. To augment the dataset labels provided by at least one expert radiologist, we recruited two more radiologists with at least ten years of experience to annotate 92 selected examples and sought unanimous agreement for usable labels. After a follow-up meeting with these two labelers to discuss carefully chosen examples upon which they and the original labeler did not all agree, we still had very low agreement on labels. Suspecting that additional labelers were more likely to agree on labels for the already agreed-upon examples, we recruited two more radiologists with at least ten years of experience to label 37 examples and found that there indeed was much higher agreement. We removed the six examples upon which there was disagreement as well as two more that were paired with one of those examples for treatment assignment. Lastly, we created six synthetic examples from agreed-upon examples by changing the prediction from "pneumonia" to "no pneumonia" or vice versa. These were added to represent scenarios for which we did not otherwise have agreed-upon examples. In the end, we had 35 examples to utilize.

\subsubsection{Example Selection and Presentation}
\label{sec:examples2}
The same pre-trained model \cite{cohen2021torchxrayvision,cohen2020limits} and dataset \cite{shih2019augmenting} used in Study 1 were used to generate examples for Study 2. This dataset's possible labels are "pneumonia," "normal," and "lung opacity, abnormal." For simplicity, we selected images labeled "normal" rather than "lung opacity, abnormal" for our "no-pneumonia" examples, except in the case of high-confidence incorrect pneumonia predictions, for which we had no other options. After finding in Study 1 that false positives seemed more preferable than false negatives in identifying pneumonia, we set the threshold at 0.56 for determining each image's classification, prioritizing sensitivity (90.2\%) over specificity (70.0\%) without decreasing specificity too much. The overall model accuracy was 74.5\%. The saliency-map explanations were again generated as Grad-CAM++ explanations \cite{chattopadhay2018grad} with PyTorch implementations \cite{jacobgilpytorchcam,simplepytorchcam}.

Each participant encountered an alert treatment and a no-alert treatment. Per participant, there were three training examples as well as nine examples per treatment. The original confidence for each example based on the 0.56 threshold was scaled to be presented between 50\% and 100\%. This scaled confidence is the confidence we will refer to moving forward. The training examples as well as five examples in each treatment had a non-anomalous confidence ranging from 69\% to 87\%, with the exception of a training example that had a 58\% confidence. The training examples consisted of one true positive, one true negative, and one false negative. The non-anomalous examples per treatment consisted of four true negatives and one true positive. Each treatment also had two low-confidence and two high-confidence examples. For each of these pairs, one was correct and one was incorrect. Also, considering the model's confidence distribution for its training data, one had a confidence within the distribution's fifth percentile (very-high and very-low) and the other within the twentieth percentile (high and low). The high and very-high confidences were 96\% and 99\% respectively. The low and very-low confidences were 51\% and 56\% respectively. For an anomalous example slot with a given accuracy and confidence category, the particular example was randomly chosen from the available relevant examples, which could have a prediction of "pneumonia" or "no pneumonia." However, given the examples available, only the very-high-confidence and very-low-confidence examples could have a "pneumonia" prediction.

For each participant, each example in the alert treatment had a matching example in the no-alert treatment with respect to confidence category (normal, very-high, high, very-low, low) and prediction category (true positive, true negative, false positive, false negative). The three training examples were randomly ordered, except the incorrect one was always second. For each treatment, the anomalous examples were randomly ordered and then assigned to positions 3, 5, 7, and 8. The non-anomalous examples were also randomly ordered and filled the remaining positions. For each matching pair correctly predicting pneumonia, we made sure that the pathology masks were comparable in size.

\subsubsection{Participants and Procedure}
\label{sec:par2}
Thirty-three radiologists recruited by email from various institutions participated in Study 2 and were compensated with a \$25 Amazon gift card. All evaluated chest x-rays at least monthly (21 daily, 10 weekly, 2 monthly), and all but one had at least a year of experience in evaluating chest x-rays (<1 year: 1, 1-5 years: 13, 6-10 years: 6, >10 years: 13). Participants reported diverse sub-specialties such as cardiothoracic radiology, pediatric radiology, neuroradiology, and interventional radiology. Their demographic distribution is as follows: 21 men, 10 women, 2 unreported; 22 white, 7 Asian, 1 Native Hawaiian or Other Pacific Islander, 1 mixed race, 2 unreported; 4 Hispanic/Latinx, 25 not, 4 unreported. Also, the participants' reported ages ranged from late twenties to late sixties. 

\begin{figure}[tb]
  \centering
  \fbox{\includegraphics[width=4cm]{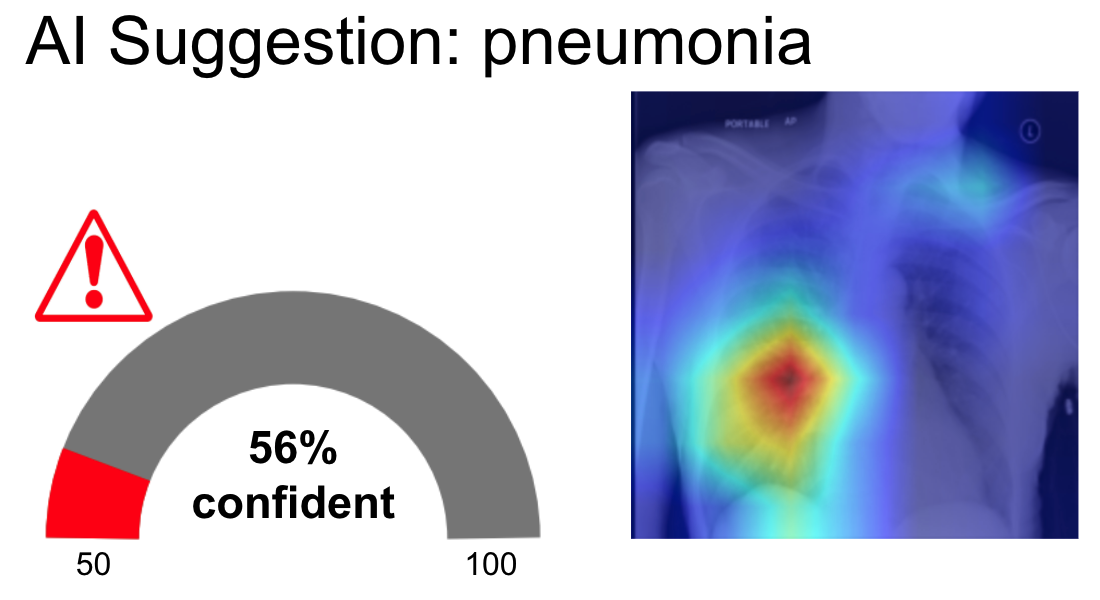}}
  \fbox{\includegraphics[width=4cm]{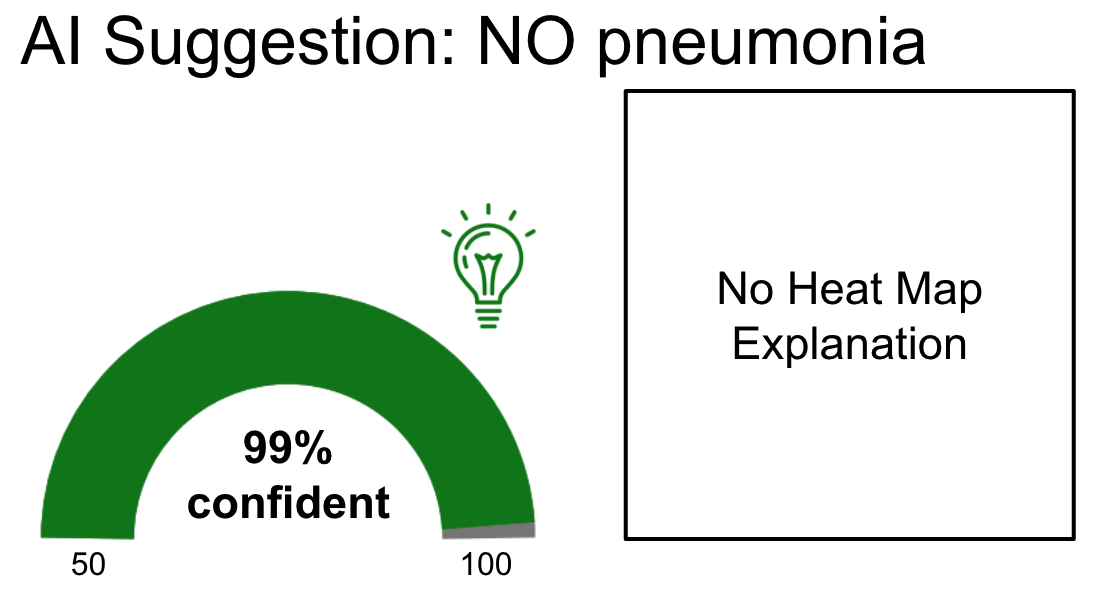}}
  \fbox{\includegraphics[width=4cm]{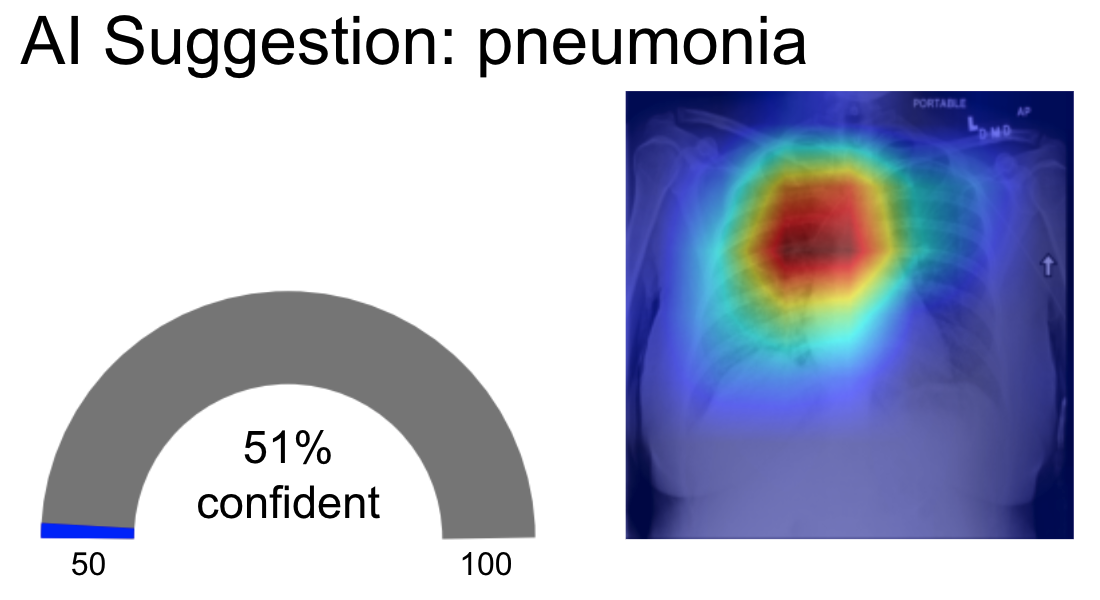}}
  \vspace*{-.1in}
  \caption{Study 2 AI suggestions with a low-confidence alert, a high-confidence alert, and a baseline no-alert despite low confidence.}
  \label{fig:alerts2}
  \Description{Each suggestion says "AI suggestion: " at the top, followed by the prediction of "pneumonia" for the low-confidence alert and baseline no-alert and "NO pneumonia" for the high-confidence alert. Underneath, the low-confidence alert and baseline no-alert have saliency-map explanations on the right, while the high-confidence alert has a box that says "No Heat Map Explanation." On the left of each suggestion is the confidence visualization, which is a semi-circle arc filled only partially to indicate the AI's confidence. Under the left foot of the arc, it says 50, and under the right foot it says 100. For the low-confidence alert, the words "56\% confident" are bolded in the middle of the arc, the partial filling of the arc is in red, and a red caution icon appears above the arc's left. For the high-confidence alert, the words "99\% confident" are bolded in the middle of the arc, the partial filling of the arc is in green, and a green light-bulb icon appears above the arc's right. For the suggestion without an alert, the words "51\% confident" are not bolded in the middle of the arc, and the partial filling of the arc is in blue.}
\end{figure}

Participants were sent a Google Forms survey\footnote{An example of the survey may be found at \href{https://forms.gle/3WbbKVnDxsc2j7t69}{this link}.} and given about a week to complete it in one sitting. Three pilot runs suggested that the completion time was approximately 10 to 20 minutes. The survey began with a study description and, to encourage participants to do their best, noted that they would later be notified of the average participant's and their survey score. The survey then moved on to a tutorial explaining the participant's task and how to interpret the AI's output. Participants were told that they would evaluate two sets of nine chest x-rays for pneumonia, and each set would be accompanied by a different AI assistant - one with special alerts and one without. As in Study 1, we explained that a confidence of 50 indicates that the AI is completely unsure if pneumonia is present or not, while a confidence of 100 indicates the opposite. We also asked participants to assume all patients have a cough and fever, as was the assumption used in labeling the full dataset \cite{shih2019augmenting}. Participants were given three attention-check questions. If they did not get all three correct, their results were removed from the study. Two were removed in this way, leaving 33 participants.

Participants then moved on to evaluate 21 chest x-rays for pneumonia. The first three were training examples for which the answers were revealed to the participants. The subsequent nine comprised the first treatment and the remaining nine the second treatment. Whether the first treatment was the alert or no-alert treatment was randomized. The x-rays allocated to each treatment are described in Section \ref{sec:examples2}. For each x-ray, the participants were first asked to evaluate it alone. They were provided with the x-ray itself as well as two links that allowed them to zoom in on the same x-ray and color-inverted version of it. They were asked to rate their agreement with the 7-point Likert-type statement "This patient has a lung opacity suspicious for pneumonia." This was asked instead of a binary question like in Study 1 because Study 1 participants noted that they usually provide a differential diagnosis rather than a binary answer when evaluating x-rays for pneumonia, similarly noted in Bussone et al. \cite{bussone2015role}. Next, they moved on to a page that looked identical except that it included the AI suggestion. The AI suggestion showed the AI's prediction, confidence, and saliency-map explanation if the prediction was "pneumonia" (Figure \ref{fig:alerts2}). Adapted from Kocielnik et al. \cite{kocielnik2019will}, a confidence visualization was provided in Study 2 to better help people comprehend the confidence. Compared to Study 1's confidence alerts, Study 2's confidence alerts were presented in a more concise manner in order to reduce cognitive load.
At the end of each treatment, participants rated their agreement with the 7-point Likert-type statement "I would use AI Assistant 1 [or 2] if it were available to me," adapted from Kocielnik et al. \cite{kocielnik2019will}.

\subsection{Results and Discussion}
To evaluate each hypothesis described in Section \ref{sec:hypo2}, we employed the Wilcoxon signed-rank test to compare within-subjects how much closer participants got to the correct answer with versus without the alert. For example, if a participant initially gave a patient a 2 on the 7-point scale for suspicion for pneumonia and after seeing the AI suggestion gave the patient a 3, the participant got one point closer to the correct answer. Also, we considered any answer on the correct side of the Likert-type scale as correct, with 4 always incorrect.
\textbf{We did not find evidence that high- or low-confidence alerts significantly impact human-AI team accuracy, whether or not the AI is correct.} The Bonferroni-corrected results did not indicate a significant difference due to alerts for correct high-confidence (V=6, p=0.69), incorrect high-confidence (V=3, p=0.93), incorrect low-confidence (V=7.5, p=1.0), or correct low-confidence (V=10.5, p=1.0) suggestions. One reason the alerts may not have had a significant effect is that the flagged confidences were very high and low in absolute terms. Had a high-confidence alert been needed for a confidence as low as 78\%, for instance, participants might have gained more from a high-confidence alert. Future work may investigate if models with larger variances in their confidence distributions benefit more from anomalous-confidence alerts. Another reason may have to do with the selection of x-rays used in Study 2. The thorough process for obtaining their labels described in Section \ref{sec:label} only yielded examples on which four expert radiologists as well as an original expert radiologist agreed. While this process ensured that the labels were highly reliable, they were likely easier to agree upon because the associated x-rays themselves were easier to evaluate.
Indeed, on average, participants provided a correct initial answer for 7 of 8 anomalous-confidence examples.
In order to better understand if confidence alerts can impact human-AI team accuracy, future work may specifically utilize data points that experts agree upon but consider difficult to evaluate.
Finally, Study 2's alerts were presented more concisely and thus perhaps more subtly than in Study 1, so they may have impacted the user less.

We used a Wilcoxon signed-rank test to compare participants' Likert-type responses regarding whether or not they would want to use the two AI assistants. \textbf{Participants did not indicate a significant difference in preference between the AI assistant with anomalous-confidence alerts versus the one without them} (V=24, p=0.44), though the alerts assistant had a slightly higher median score (4 vs 3). This contradicts the overall positive reaction to these alerts in Study 1. That said, this negative result may have been influenced by the same points described above in relation to the hypothesis results. 
Furthermore, unlike in Study 1, the preference question conflated high- and low-confidence alerts; radiologists may still desire high-confidence alerts more than low-confidence ones.

\section{Limitations}
Given the specialized participant pool, both studies had a limited number of participants. Also, neither study was longitudinal, which would provide better insight into how the proposed interventions would work in practice. Although Study 2 provided some abilities that make evaluating x-rays easier, the studies were conducted outside of physicians' normal environment for evaluating x-rays. Unlike in reality, participants were not provided additional information regarding each patient and were not allowed to consider diagnoses other than pneumonia. In addition, a limited number of example x-rays were used in each study and may not be representative of x-rays evaluated for pneumonia at large. Though based upon real situations, the anomalous input and explanation examples were hand-picked rather than automatically detected. Finally, Study 1's anomalous scenarios were likely presented with unrealistically high frequency.

\section{Conclusion}
We explored how alerts for anomalous model input, confidence, and saliency-map explanations may impact users of a CDSS used to evaluate x-rays for pneumonia. In a formative study, we interviewed and surveyed 4 radiologists and 4 other physicians about their interactions with mockups of CDSS anomaly alerts. We found evidence suggesting that non-radiologist physicians who regularly evaluate chest x-rays desire the four proposed AI anomaly alerts, and high-confidence alerts are desirable among both radiologists and other physicians. In a follow-up user study, 33 radiologists engaged with two CDSS treatments, one with and one without high- and low-confidence alerts. We did not observe evidence indicating that these alerts improved radiologist-AI team accuracy or radiologists' experience. 
Future work may continue to explore if and how AI suggestions should be provided in different anomalous cases. 
We hope that this work acts as a building block towards more research on helping users manage anomalous model input and output.

\begin{acks}
This research was supported by Microsoft, NSF RAPID grant 2040196, ONR grant N00014-18-1-2193, and the Allen Institute for Artificial Intelligence (AI2). The authors thank the many people who provided helpful feedback on the project, in particular Rashmi Raj, Steven Borg, Matthew Lungren, and Ozan Oktay. The authors also thank the participants who made this work possible and the anonymous reviewers of this work for their valuable feedback.
\end{acks}

\bibliographystyle{ACM-Reference-Format}
\bibliography{sample-base}

\end{document}